\documentclass[aps,pra,superscriptaddress,showpacs,showkeys]{revtex4}
\usepackage{epsfig}
\usepackage{graphicx}
\usepackage{bm}
\usepackage{amssymb}
\newcommand{\zh}{\bm}
\newcommand{\real}{\mathop{\rm Re}\nolimits}

\newcommand{\mee}{{E}}

\newcommand{\zhr}{{\zh r}}

\newcommand{\zhe}{{\zh e}}

\newcommand{\zhp}{{\zh p}}
\newcommand{\zhk}{{\zh k}}

\newcommand{\zhA}{{\zh A}}
\newcommand{\zhY}{{\zh Y}}

\newcommand{\Eq}[1]{Eq.\ (\ref{#1})}

\newcommand{\txt}[1]{{\rm #1}}

\newcommand{\vectorc}[3]{
\left(
\begin{array}{c}
#1\\
#2\\
#3\\
\end{array}
\right)
        }

\begin{document}
\title{Calculation of differential cross section for dielectronic recombination with two-electron uranium}

\author{K.\ N.\ Lyashchenko}
\email{laywer92@mail.ru}
\affiliation{Department of Physics,
             St.\ Petersburg State University, 7/9 Universitetskaya nab.,
             St. Petersburg,
             199034,
             Russia}
\affiliation{ITMO University,
             Kronverkskii ave. 49, 197101,
             Petergof, St.\ Petersburg, Russia}
\author{O.\ Yu.\ Andreev}
\affiliation{Department of Physics,
             St.\ Petersburg State University, 7/9 Universitetskaya nab.,
             St. Petersburg,
             199034,
             Russia}

\date{\today}

\begin{abstract}
Calculation of the differential cross section for the dielectronic recombination with two-electron uranium within the framework of QED is presented.
The polarization of the emitted photon is investigated.
The contributions of the Breit interaction and the interference of the photon multipoles are studied.
\end{abstract}

\pacs{31.10.+z, 31.15.ac, 31.30.J-, 34.70.+e}
\keywords{electron recombination, dielectronic recombination, QED, highly charged ions, three-electron ions}

\maketitle

%
\section{Introduction}
Dielectronic recombination with few-electron (in particular, with two-electron) highly charged ions is of interest both from experimental and theoretical points of view.
Few-electron highly charged ions are relatively simple systems which allow precise theoretical description within the framework of quantum electrodynamics (QED).
Dielectronic recombination is a resonant process where electron recombination with an ion is performed via formation of an autoionizing (doubly excited) state of the ion.
Interelectron interaction plays a crucial role in formation of the doubly excited states.
Accordingly, dielctronic recombination presents a tool for detailed investigation of dynamic electron correlations, in particular, for investigation of the Breit interaction.

Dielectronic recombination with two-electron ion of Fe was studied experimentally and theoretically in work
\cite{beiersdorfer92}.
Measurements of the radiative and Auger decay rates for K-shell vacancies in highly charged Fe ions were presented in
\cite{steinbrugge2015}.

The linear polarization measurements of x-rays emitted due to dielectronic recombination into highly charged Kr ions were recently presented in
\cite{shah2015}.

Measurements of the dielectronic recombination resonant strengths of highly charged ions (in particular, two-electron) Xe ions were performed in work
\cite{yao2010}.

Experimental investigation of dielectronic recombination with two- and three-electron ions of Pr, Ho and Au is reported in
\cite{hu2014}.
In particular,  dielectronic recombination strengths are measured and calculated.
Experimental study of the dielectronic recombination with three-electron U ions is presented in
\cite{trotsenko2015}.

Calculation of the transition rates for the doubly excited states of a three-electron ion of uranium was reported in
\cite{natarajan2015}.
The influence of the Breit and QED effects on the radiative transition parameters is analyzed in detail.
Dielectronic recombination with two-electron uranium ion was also studied theoretically in 
\cite{chen2015}
In particular, the linear polarization and angular distribution of the x-ray photoemission was studied,  contribution of the magnetic quadrupoles was investigated.

Dielectronic recombination with one-electron uranium was studied experimentally and theoretically.
The experimental study of the full cross section of was performed in
\cite{bernhardt11}.
The corresponding calculations of the dielectronic recombination are presented in
\cite{karasiov92p453,zakowicz04,andreev09p042514,bernhardt11,lyashchenko15}.

\par
The electron recombination with highly charged ions presents a tool for investigation of the Breit interaction.
The recent studies showed that the Breit interaction may give important and even dominant contribution to the cross section of the dielectronic recombination  with few-electron highly charged ions
\cite{lyashchenko15,Nakamura2008PhysRevLett.100.073203,Fritzsche2009PhysRevLett.103.113001,Matula2011PhysRevA.84.052723,bernhardt11,hu12,shah2014hci}.

\par
The considered process of electron recombination with a two-electron uranium can be described as
\begin{eqnarray}
e^{-}+U^{90+}(1s1s)
&\to&\label{eq080821n01}
U^{89+}(r)+\gamma
\,,
\end{eqnarray}
where $r$ denotes a singly excited state: $(1s1s2s)$ or $(1s1s2p)$.
In the initial state of the system the two-electron ion is assumed to be in its ground state.
The first emitted photon ($\gamma$) can be registered in the experiment.
If the energy of the initial state is close to the energy of a doubly excited state, the cross section shows resonances.
The resonances are explained by the dielectronic recombination which can be written as
\begin{eqnarray}
e^{-}+U^{90+}(1s1s)
&\to&\label{eq080821n02}
U^{89+}(d)
\to
U^{89+}(r)+\gamma
\to\ldots
\,,
\end{eqnarray}
where $d$ is a doubly excited state: $(1s2s2s)$,  $(1s2s2p)$ or  $(1s2p2p)$.

To study the cross section of the dielectronic recombination with highly charged ions, the QED calculations of the radiative transitions amplitudes between three-electron configurations are necessary.
Such calculations can be performed with employment of various methods
\cite{johnson95,shabaev02,indelicato04,andreev08pr,andreev09p032515,andreev09p042514}.
In the present paper the line-profile approach was used
\cite{andreev08pr}.

\par
We present calculations of the full and differential cross section for the dielectronic recombination with two-electron uranium within the framework of QED.
The polarization of the emitted photon is investigated.
The contributions of the Breit interaction and the interference of the photon multipoles are studied.
At the end, we will estimate the contribution of the three electron recombination for the considered collision system.

%
%
%
\section{Method of calculation}
The present calculations are based on the QED approach already applied for calculation of the cross section of electron recombination with one-electron ions
\cite{andreev09p042514,lyashchenko15}
-- the line-profile approach (LPA)
\cite{andreev08pr}.
To describe the electron recombination with two-electron ions, the line-profile approach was generalized to three-electron system.

The incident electron is considered as an electron with certain momentum $\zhp$ and polarization $\mu$ and is described by wave function $\psi_{\zhp,\mu}$.
Wave function $\psi_{njlm}$ describes a bound electron with the corresponding quantum numbers ($n$,$j$,$l$ and $m$ are the principle quantum number, total angular momentum, parity and projection of $j$, respectively).
Employing expansion of the wave function $\psi_{\zhp,\mu}$ in series over the wave functions of electron with certain energy ($\varepsilon$), total angular momentum ($j$), its projection ($m$), and parity ($l$)
the wave function of the incident electron can be written as
\cite{akhiezer65b}
\begin{eqnarray}
\psi_{\zhp, \mu}(\zhr)
&=&\label{psie}
\int d\varepsilon \sum_{jlm} a_{{\zhp} \mu, \varepsilon jlm}\psi_{\varepsilon jlm}(\zhr)
\,,
\\
a_{{\zhp} \mu, \varepsilon jlm}
&=&\label{coefa}
\frac{(2\pi)^{3/2}}{\sqrt{p\epsilon}}i^l e^{i\phi_{jl}}
(\Omega^+_{jlm}(\zhp), \upsilon_{\mu}(\zhp))
\delta(\epsilon-\varepsilon)
\,,
 \end{eqnarray}
where $\Omega_{jlm}(\zhp)$ is the spherical spinor and $\upsilon_{\mu}(\zhp)$ is the
spinor with certain projection on the electron momentum ($\zhp$), the phase $\phi_{jl}$ is the Coulomb phase shift.
The relativistic units are used throughout unless otherwise stated.

In the line-profile approach the initial, final and intermediate three-electron states are expanded is series of three-electron functions in the $j$-$j$ coupling scheme.
The following algorithm is employed for composition of the basis set of three-electron wave functions in the $j$-$j$ coupling scheme.
Firstly, we select two electrons with closest energies and a compose two-electron wave function in the $j$-$j$ coupling scheme
\begin{eqnarray}
\Psi^{(0)}_{j_{12}m_{12}n_1j_1l_1n_2j_2l_2}
&=&\label{eqn151216n02}
N
\sum\limits_{m_1m_2}
C^{j_1j_2}_{j_{12} m_{12}}(m_1m_2)\det\{\psi_{n_1j_1l_1m_1},\psi_{n_2j_2l_2m_2}\}
\,,
\end{eqnarray}
where $N$ -- normalization constant (equal to $1/\sqrt{2}$ for nonequivalent electrons and $1/2$ for equivalent electrons),
$C^{j_{1}j_2}_{j_{12}m_{12}}(m_{1}m_2)$ -- Clebsch-Gordan coefficients.
The electrons are represented by the quantum numbers $n_i$,$j_i$,$l_i$ and $m_i$ ($n_i$ denotes the principle quantum number for bound electron and the energy for electron from the continuum part of the spectrum, $j_i$ is the total angular momentum, $l_i$ is the parity, $m_i$ is the projection of the total angular momentum, $i=1,2$ denotes the first and second electrons, respectively).
Then, we compose three-electron wave functions with certain angular momentum ($J$) and its projection $M$
\begin{eqnarray}
\Psi^{(0)}_{JMj_{12}n_1j_1l_1n_2j_2l_2 n_3 j_3l_3}
&=&\label{eqn151216n03}
N
\sum\limits_{m_1m_2m_{12}m_3}
C^{j_{12}j_3}_{JM}(m_{12}m_3)C^{j_1j_2}_{j_{12} m_{12}}(m_1m_2)\det\{\psi_{n_1j_1l_1m_1},\psi_{n_2j_2l_2m_2},\psi_{n_3 j_3l_3m_3}\}
\,,
\end{eqnarray}
where $N$ -- normalization constant (equal to $1/\sqrt{ 3!\, 2}$ for tree-electron states with two equivalent electrons and to $1/\sqrt{3!}$ for states without equivalent electrons).
If three-electron configurations contain three equivalent electrons, the coefficients of fractional parentage ($ \langle j_1j_2 [j_{12}]j_3 J\} j_1j_2j_3\gamma J \rangle $) are to be calculated for composing the three-electron wave functions in  $j$-$j$ coupling scheme
\cite{sobelman63,veselov86b}
\begin{eqnarray}
\Psi^{(0)}_{JM\gamma n_1j_1l_1n_2j_2l_2n_3j_3l_3}
&=&\label{eqn151225n01}
\sum\limits_{j_{12}}
\langle j_1j_2 [j_{12}]j_3 J\} j_1j_2j_3\gamma J \rangle\Psi^{(0)}_{JMj_{12}n_1j_1l_1n_2j_2l_2n_3j_3l_3}
\,.
\end{eqnarray}
Here, the quantum number $\gamma$ denotes repeating terms of the electronic structure (if necessary).

The initial state of the system (a two-electron ion in its ground $(1s1s)$ state and an incident electron) can be described by the wave function
\begin{eqnarray}
\Psi^{\txt{ini}}
&=&\label{ininonint}
\frac{1}{\sqrt{3!}}\det\{\psi_{n_1j_1l_1m_1},\psi_{n_2j_2l_2m_2},\psi_{\zhp,\mu}\}
\,.
\end{eqnarray}
The wave function $\psi_{\zhp,\mu}$ describes the incident electron with momentum $\zhp$ and polarization $\mu$, the wave function $\psi_{njlm}$ describes a bound electron with the corresponding quantum numbers ($n$,$j$,$l$ and $m$).
Employing expansion
\Eq{psie}
the wave function of the initial state can be presented as a linear combination of the following determinants
\begin{eqnarray}
\Psi^{(0)}_{n_1j_1l_1m_1n_2j_2l_2 m_2\varepsilon j_3l_3m_3}
&=&\label{eqn151225n02}
\frac{1}{\sqrt{3!}}
\det\{\psi_{n_1j_1l_1m_1},\psi_{n_2j_2l_2 m_2},\psi_{\varepsilon j_3l_3m_3}\}
\,.
\end{eqnarray}
Accordingly, the wave function of the initial state can be expanded into series of three-electron functions with certain total momentum (in the $j$-$j$ coupling scheme)
\begin{eqnarray}
\Psi^{\txt{ini}}
&=&\label{psiini}
\sum\limits_{JMj_{12}n_1j_1l_1n_2j_2l_2 j_3l_3}
\int d\varepsilon\,
\langle \Psi^{(0)}_{JMj_{12}n_1j_1l_1n_2j_2l_2\varepsilon j_3l_3}\,|\,\Psi^{\txt{ini}}\rangle\,
\Psi^{(0)}_{JMj_{12}n_1j_1l_1n_2j_2l_2\varepsilon j_3l_3}
\,.
\end{eqnarray}

The final state ($(1s1s2s)$, $(1s1s2p)$ tree-electron states) can be written as a tree-electron configuration in the $j$-$j$ coupling scheme
\begin{eqnarray}
\Psi^{\txt{fin}}
&=&\label{psifin}
\Psi^{(0)}_{JMj_{12}n_1j_1l_1n_2j_2l_2n_3j_3l_3}
\,.
\end{eqnarray}


\par
To describe highly charged ion within the framework of QED, the line-profile approach (LPA) was employed
\cite{andreev08pr,andreev09p042514}.
The interaction with the quantized electromagnetic and electron-positron fields leads to various correction to the amplitude: interelectron interaction correction, electron self-energy and vacuum polarization corrections. 
Within the line-profile approach the system is considered to be enclosed into a sphere of a large radius $R\to\infty$.
Then, the incident electron can be described by a wave function normalized to unit which corresponds to a artificial bound electron state ($e_R$).
For calculation of the amplitudes of the transitions between bound states the standard QED perturbation theory for the quasidegenerate states can be applied
\cite{shabaev02,lindgren04,Lindgren2014PhysRevA.89.062504,andreev08pr}.

Within the LPA we introduce the set of three-electron configurations ($g$) which includes all the three-electron configurations composed by
$1s,2s,2p,3s,3p,3d$ electrons and the electron $e_R$ describing the incident electron.
There is also introduced the matrix $V$ which is defined by the one and two-photon exchange, electron self-energy and vacuum polarization matrix elements.
Matrix $V=V^{(0)}+\Delta V$ is considered as a block matrix
\begin{eqnarray}
V 
&=&\label{V_mat}
        \left[
        \begin{array}{cc}
        V_{11} & V_{12}  \\
        V_{21}  &V_{22}  \\
        \end{array}
        \right]
\,=\,
        \left[
        \begin{array}{cc}
       V^{(0)}_{11}+ \Delta V_{11} & \Delta V_{12}  \\
        \Delta V_{21}  &V^{(0)}_{22}+\Delta V_{22}  \\
        \end{array}
        \right]
\,.
\end{eqnarray}
Matrix $V_{11}$ is defined on set
$g$,
which contains configurations mixing with the reference state
$n_g$ $\in$ $g$.
Matrix $V^{(0)}$ is a diagonal matrix: sum of the one-electron Dirac energies.
Matrix $\Delta V_{11}$ is not a diagonal matrix, but it contains a small parameter of the QED perturbation theory.
Matrix $V_{11}$ is a finite-dimensional matrix and can be diagonalized numerically.
Then, the standard perturbation theory can be applied for the diagonalization of the matrix $V$.

\par
The amplitude of the electron recombination is written as a matrix element of the photon emission operator ($\Xi^{(0)}$) with the bra and ket vectors given by the eigenfunctions of the matrix $V$:
$\Psi^\txt{fin}$ and $\Psi^\txt{ini}$ corresponding to the final and initial states of the system, respectively,
\begin{eqnarray}
U_{if}
&=&
\langle\Psi^\txt{fin}|\Xi|\Psi^\txt{ini}\rangle
\,.
\end{eqnarray}
The operator $\Xi$ is derived within the QED perturbation theory order by order
\cite{andreev08pr,andreev09p042514}.
The operator $\Xi$ can be represented by its matrix elements, in the zeroth order of the perturbation theory it reads
\begin{eqnarray}
\Xi^{(0)}_{u_1u_2u_3d_1d_2d_3}
&=&
eA^{(k,\lambda)*}_{u_1d_1}
\delta_{u_2d_2}\delta_{u_3d_3}
+
\delta_{u_1d_1}eA^{(k,\lambda)*}_{u_2d_2}
\delta_{u_3d_3}
+
\delta_{u_1d_1}\delta_{u_2d_2}eA^{(k,\lambda)*}_{u_3d_3}
\,,
\end{eqnarray}
where $u_1$, $u_2$, $u_3$, $d_1$, $d_2$, $d_3$ are one-electron states with certain total angular momentum and parity, the one-electron matrix elements $A^{(k,\lambda)*}_{ud}$ are defined as
\begin{eqnarray}
A^{(k,\lambda)*}_{ud}
&=&
\int d^{3} {\zhr} \,
\overline{\psi}_{u}({\zhr})\gamma^{\nu}A^{(k,\lambda)*}_{\nu}(\zhr)
\psi_{d}(\zhr)
\,,
\end{eqnarray}
where $\gamma^{\nu}$ are Dirac gamma matrices. $A^{(k,\lambda)*}_{\nu}=(A_0^{(k,\lambda)*}, \zhA^{(k,\lambda)*})$ is the emitted photon wave function. We use a gauge in which $A_0^{(k,\lambda)}=0$.  $\zhA^{(k,\lambda)}$ reads as
\begin{eqnarray}
{\zhA}^{(k,\lambda)}({\zhr})
&=&
\sqrt{\frac{2\pi}{\omega}} e^{i \zhk\zhr} \zhe^{(\lambda)}
\,,
\end{eqnarray}
$\omega$ and $ \zhk$ are frequency and momentum of the photon, respectively.
Employing the multipole expansion we can write
\cite{labzowsky96b}
\begin{eqnarray}
 {\zhA}^{(k,\lambda)}
&=&\label{12345}
\sqrt{\frac{2\pi}{\omega}}
\sum_{j_0l_0m_0}i^{l_0} g_{l_0}(\omega r)
(\zhe^{(\lambda)}, {\zhY}_{j_0l_0m_0}^*(\zhk)) \zhY_{j_0l_0m_0}(\zhr)
\,,
\end{eqnarray}
where $g_{l_0}(x)=4\pi j_{l_0}(x)$ and $j_{l_0}(x)$ is the spherical Bessel function,
$\zhY_{j_0l_0m_0}$ -- vector spherical harmonics, $\zhe^{(\lambda)}$ -- vector of photon polarization.
We consider the linear polarization of the photon
\begin{eqnarray}
\zhe_1
&=&
\frac{[\zhp\times\zhk]}{|[\zhp\times\zhk]|}
\,,\qquad
\zhe_2
\,=\,
\frac{[\zhe_1\times\zhk]}{|[\zhe_1\times\zhk]|}
\end{eqnarray}
and the circular polarization of the photon
\begin{eqnarray}
\zhe_{+}
&=&\label{cirpol}
\frac{1}{\sqrt{2}}
(\zhe_1+i\zhe_2)
\,,\qquad
\zhe_{-}
\,=\,
\frac{1}{\sqrt{2}}
(\zhe_1-i\zhe_2)
\,.
\end{eqnarray}

The $z$ axis is defined by the incident electron momentum $\zhp$.
Accordingly, the  vectors $\zhp$, $\zhk$ and the polarization vectors look like
\begin{eqnarray}
\frac{\zhp}{|\zhp|}
&=&
\vectorc{0}{0}{1}
\,,\qquad
\frac{\zhk}{|\zhk|}
\,=\,
\vectorc{\sin\theta\cos\phi}{\sin\theta\sin\phi}{\cos\theta}
\,,
\\
\zhe_1
&=&\label{pol1}
\vectorc{-\sin\phi}{\cos\phi}{0}
\,,\qquad
\zhe_2
\,=\,
\vectorc{\cos\theta\cos\phi}{\cos\theta \sin\phi}{-\sin\theta}
\,,
\end{eqnarray}
respectively. In the electron-ion collision process has an axial symmetry and it not depend from angle $\phi$.

The operator $\Xi$ in the first order of the perturbation theory gives a small contribution and is omitted in the present calculation.

%
%
%
\section{Results}
We have studied the process of electron recombination with two-electron uranium being initially in its ground state.
The process is considered in the rest frame of the uranium ion.
We investigated regions of the incident electron energy where the role of dielectronic recombination is prominent.
We restricted ourselves to the consideration of four low lying energy regions, in particular, the regions where the energy of the initial state ($(1s1s)$ plus incident electron $e$) is close to the energies of doubly excited states
$(1s2s2s)$, $(1s2s2p)$, $(1s2p2p)$.
Accordingly, we performed the calculations for the following resonance regions of the incident electron (kinetic) energy: [$63.03$,$63.075$] keV, [$63.075$,$63.45$] keV, [$67.25$,$68.00$] keV and [$71.95$,$72.15$] keV.

\begin{figure}[h]
\begin{minipage}{40pc}
\includegraphics[width=40pc]{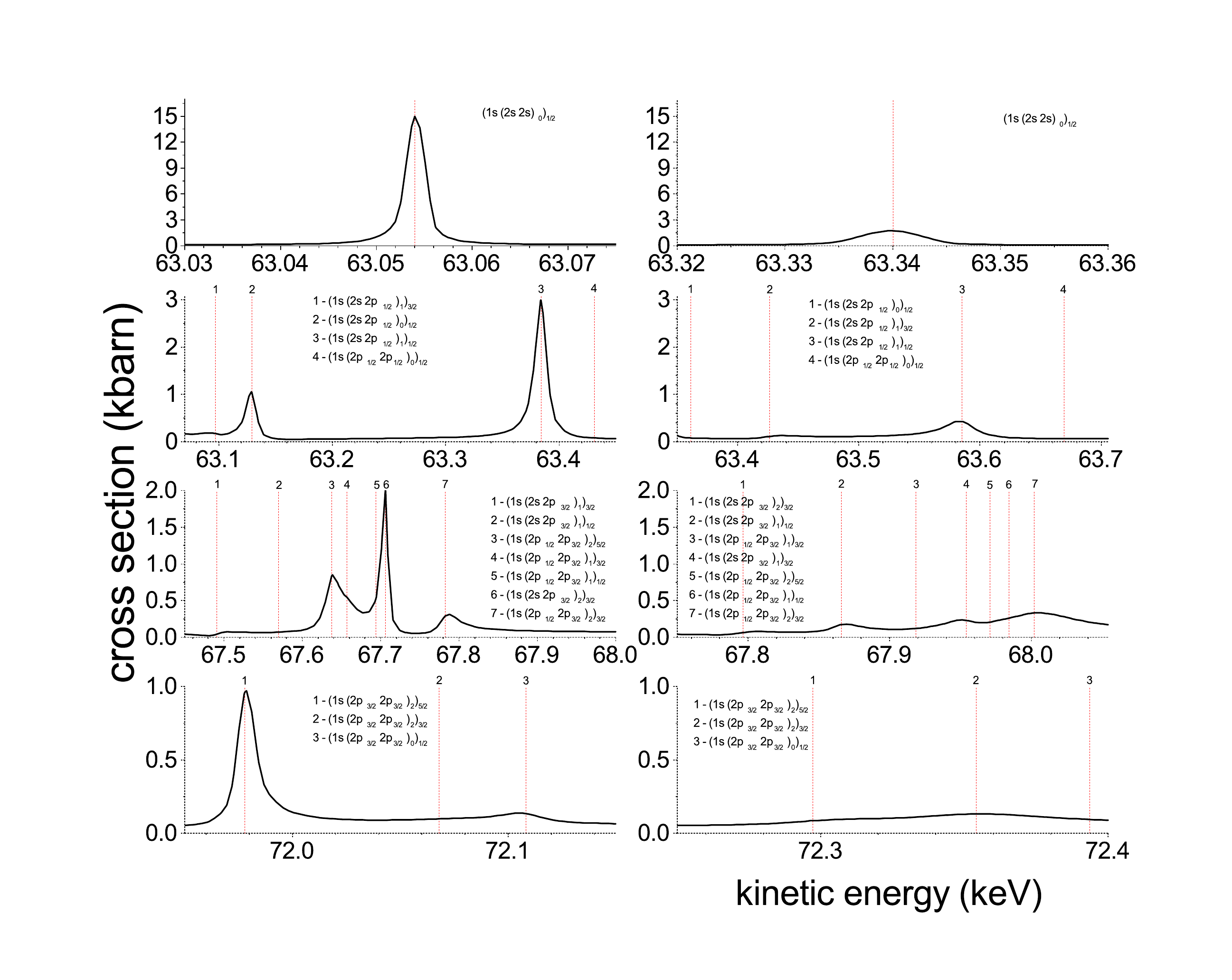}
\caption{The total cross section of the dielectronic recombination with two-electron uranium as a function of the kinetic energy of the incident electron is presented (in kbarn).
The left column correspond to full QED calculation (Coulomb + Breit with retardation) and the right column denote the calculation with disregard of the Breit interaction. Red dashed vertical lines indicate the positions of the resonances with the doubly excited states.
}
\label{fig1}
\end{minipage}\hspace{2pc}%
\end{figure}

The total cross section of electron recombination with two-electron uranium is presented as a function of the kinetic energy of the incident electron in
Fig. \ref{fig1}.
The left four graphs represent the exact QED calculation of the total cross section for the four resonance regions, respectively.
The right ones represent the calculation of the total cross section with disregard of the Breit interelectron interaction (in the Feynman graphs representing to the one- and more photon exchange).
The graphs reveal a large contribution of the Breit interaction to the cross section.

We note that the relative contribution of the Breit interaction to the cross section for dielectronic recombination with two-electron uranium ions is much larger than that for dielectronic recombination with one-electron ions (see
\cite{bernhardt11,lyashchenko15}).
The importance of the Breit interaction is explained by large sensitivity of the widths of three-electron energy levels to the Breit interaction.
Within the framework of the standard QED theory, the energy shift of energy levels (due to the interaction with quantized electromagnetic and electron-positron fields) is commonly written as
$\Delta\mee=\real\{\Delta \mee\}-i\frac{\Gamma}{2}$
\cite{low52,shabaev02,lindgren04,andreev08pr},
where $\real\{\Delta\mee\}$ is a correction to the energy,
$\Gamma$ defines the width of the energy level.
For the one- and two-electron configurations the major contribution to the width of energy level is given by the electron self-energy Feynman graph.
However, for three-electron configurations the contribution of the electron self-energy graph can be considerably canceled by the contribution of the Breit part of the one-photon exchange graph.
For example, $(1s1s2s)$ configuration is the ground state of tree-electron ion, accordingly, contribution of the imaginary part of the electron self-energy graph is completely canceled by contribution of the Breit part of the one-photon exchange graph.
It can be referred as a realization of the Pauli exclusion principle
\cite{labzowsky96b}.

The energies and widths of the considered doubly excited states are very sensitive to the Breit interaction.
In
Table~\ref{table2}
we present data which show the role of the Breit interaction for the doubly excited states.
Doubly excited states are specified in the first column.
In the second and the third columns ( '$V$ Coulomb' and '$V$ Coulomb+Breit') presented are values of the diagonal matrix elements of matrix $V$ (see
\Eq{V_mat});
these data are given by summation of the diagonal matrix elements of the corresponding Feynman graphs (electron self-energy, vacuum polarization, one-photon exchange and part of the two-photon exchange graphs).
The column '$V$ Coulomb+Breit' contains results of the exact QED calculation, the column '$V$ Coulomb' presents results of the calculation with disregard of the Breit part of photon exchange graphs.
These data have no clear physical meaning; however, they demonstrate a strong cancellation of the imaginary parts of the electron self-energy and the Breit part of the photon exchange corrections for some of the configurations.
The next three columns present results of calculation of the energies and widths of the doubly excited states performed within the exact QED approach (column 'Coulomb+Breit (with retardation)'), with disregard of retardation in the Breit interaction (column 'Coulomb+Breit (without retardation)') and with complete neglect of the Breit interaction (column 'Coulomb'). The corresponding resonance kinetic energies of the incident electron are also given.
These data demonstrate the importance of the Breit interaction for the energies and widths of the doubly excited states what explains the large contribution of the Breit interaction to the cross section for the dielectronic recombination with two-electron uranium ions which is seen in
Fig. \ref{fig1}.

\par
\begin{figure}[h]
\begin{minipage}{40pc}
\begin{center}
\includegraphics[width=40pc]{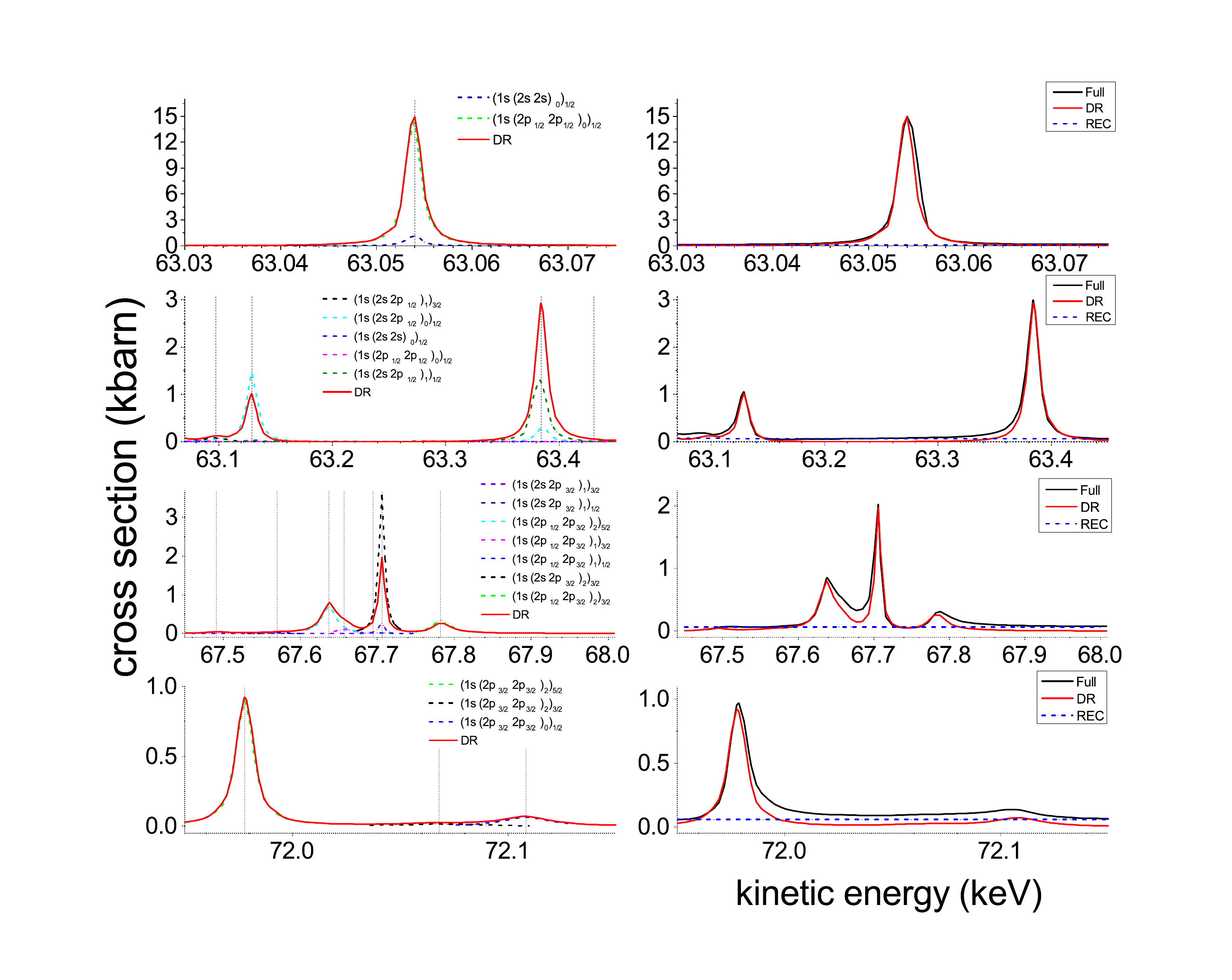}
\caption{Interference figures. The left column corresponds to the interference of the individual peaks with each other. Colorful dashed curves represent the total cross section for the individual peaks (without REC). The red solid curves represents the total cross section for dielectronic recombination (without REC). Vertical lines indicate the positions of the resonances with the doubly excited states. The right column corresponds to the interference of DR (red dashed curves) and REC (blue dots lines). Black solid curves denotes full calculation.}
\label{fig2}
\end{center}
\end{minipage}\hspace{2pc}%
\end{figure}

The process of dielectronic recombination proceeds via formation of doubly excited states.
Peaks of the cross section correspond to these doubly excited states.
In order to study the individual contributions of the doubly excited states, we performed calculations of the cross section where only fixed doubly excited states were taken into account.
The results are presented in
Fig. \ref{fig2}.
Plots in the left column show individual contribution of the doubly excited states.
Plots in the right column show separate contribution of resonant channel (electron capture via formation of doubly excited states, i.e., dielectronic recombination) and nonresonant channel -- radiative electron capture (REC).
We note that partition of the electron recombination into resonant and nonresonant channels is ambiguous.
Results of the full calculation of the cross section are given with a mark (Full) in
Fig. \ref{fig2}.
These plots also show interference between the dielectronic recombination and the radiative electron capture.

\par
\begin{figure}[h]
\begin{minipage}{40pc}
\includegraphics[width=40pc]{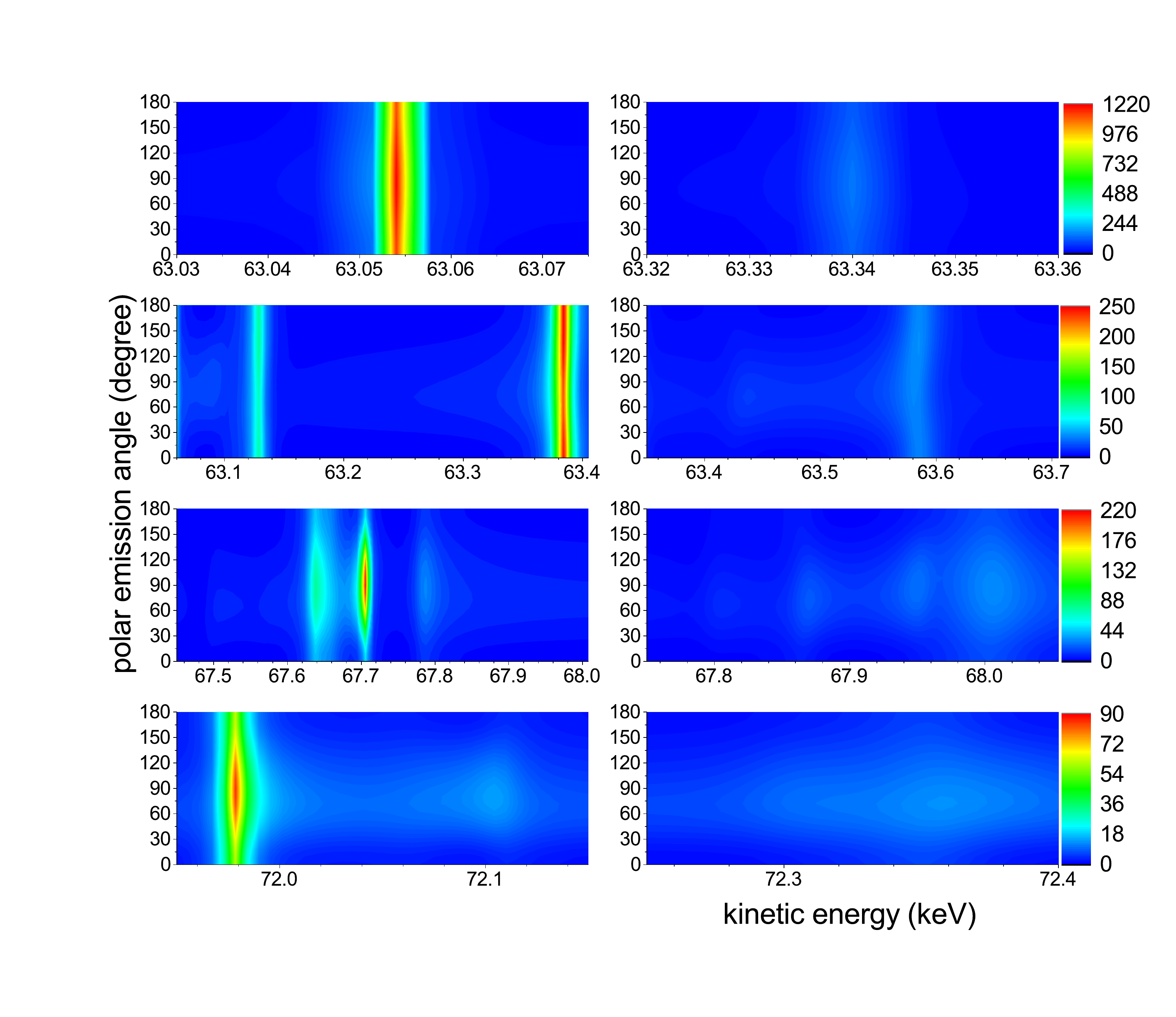}
\caption{The differential cross section ($d\sigma/d\Omega$, in barn/str) of dielectronic recombination with two-electron ions of uranium are presented.
The graphs in the left column correspond to the full QED calculation (Coulomb + Breit with retardation) of the differential cross section, the graphs in the right column present the calculation with disregard of the Breit interaction.}
\label{fig3}
\end{minipage}\hspace{2pc}%
\end{figure}

Results of calculation of the differential cross section (in barn/str)
\begin{eqnarray}
\sigma'
&\equiv&\label{eq151216n01}
d\sigma / d\Omega
\end{eqnarray}
as a function of the kinetic energy of incident electron are given in
Fig. (\ref{fig3}).
The left plots present the exact QED calculation, the right plots present results of calculation with disregard of the Breit interaction.
As a consequence of large contribution of the Breit interaction to the total cross section (see
Fig. \ref{fig1}),
the Breit interaction is also important for the differential cross section of electron capture by two-electron uranium ion.

\begin{figure}[h]
\begin{minipage}{40pc}
\includegraphics[width=40pc]{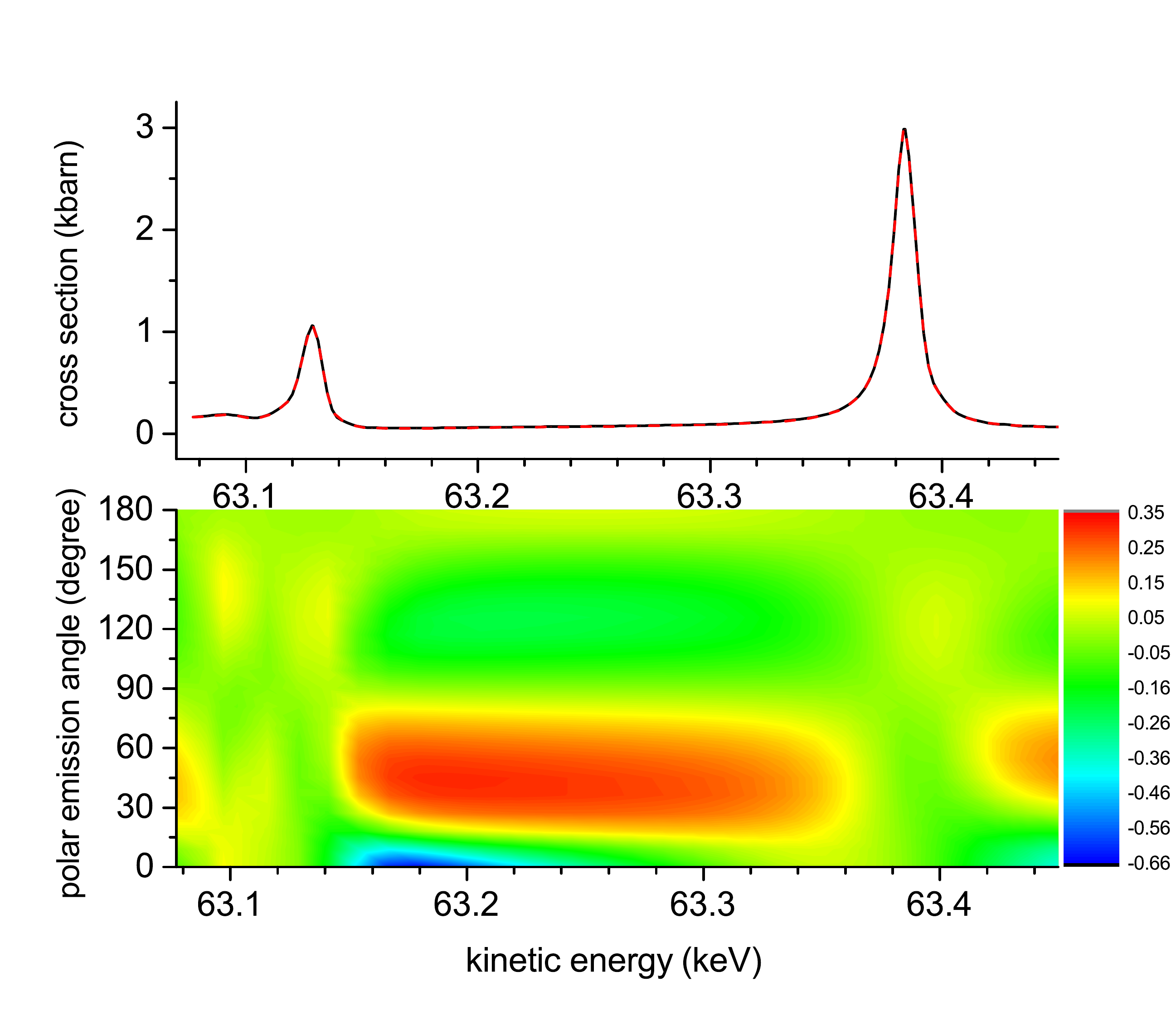}
\caption{The contribution of the higher multipoles to the full and differential cross section is presented.
The top graph correspond to full cross section (in kbarn). Red dashed curve represent calculation where just $j_0=1$ (see
\Eq{12345}) taken into account. Black solid curve denote calculation with $0<j_0<10$.  
The bottom graph represents the relative contribution of the higher multipoles
$\delta \sigma$ (see
\Eq{deltasigma}).
}
\label{fig4}
\end{minipage}\hspace{2pc}%
\end{figure}

In the present calculation the multipole expansion of the emitted photon wave function was employed (see \Eq{12345}).
The multipoles up to $j_0=9$ were taken into account.
Investigation of contribution of the higher multipoles of emitted photon is presented in
Fig. \ref{fig4}.
The upper plot presents the total cross section.
The red curve in the upper plot corresponds to calculation of the total cross section within dipole approximation where only terms with $j_0=1$ are taken into account in the multipole expansion
\Eq{12345}.
The black curve in the upper plot corresponds to the full calculation ($j_0\le 9$).
It is seen that contribution of the higher multipoles to the total cross section is insignificant.
The lower plot presents investigation of the differential cross section:
there is a relative difference between differential cross section calculated in dipole approximation
$\sigma'^{(j_0=1)}$
and the full calculation
$\sigma'$
(see
\Eq{eq151216n01})
\begin{eqnarray}
\delta\sigma
&=&\label{deltasigma}
\frac{\sigma'^{(j_0=1)} - \sigma'} {\sigma'}
\,.
\end{eqnarray}
In spite of a small contribution of the higher multipoles to total cross section ($<5 \%$), they  play a significant role for the differential cross section.
Our calculations show, that mainly due to the interference between $E1$, $M1$ and $E2$, $M2$ emitted photons, $\delta\sigma$ may reach up to $66\%$ in regions between the peaks.

\par
\begin{figure}[h]
\begin{minipage}{40pc}
\includegraphics[width=20pc]{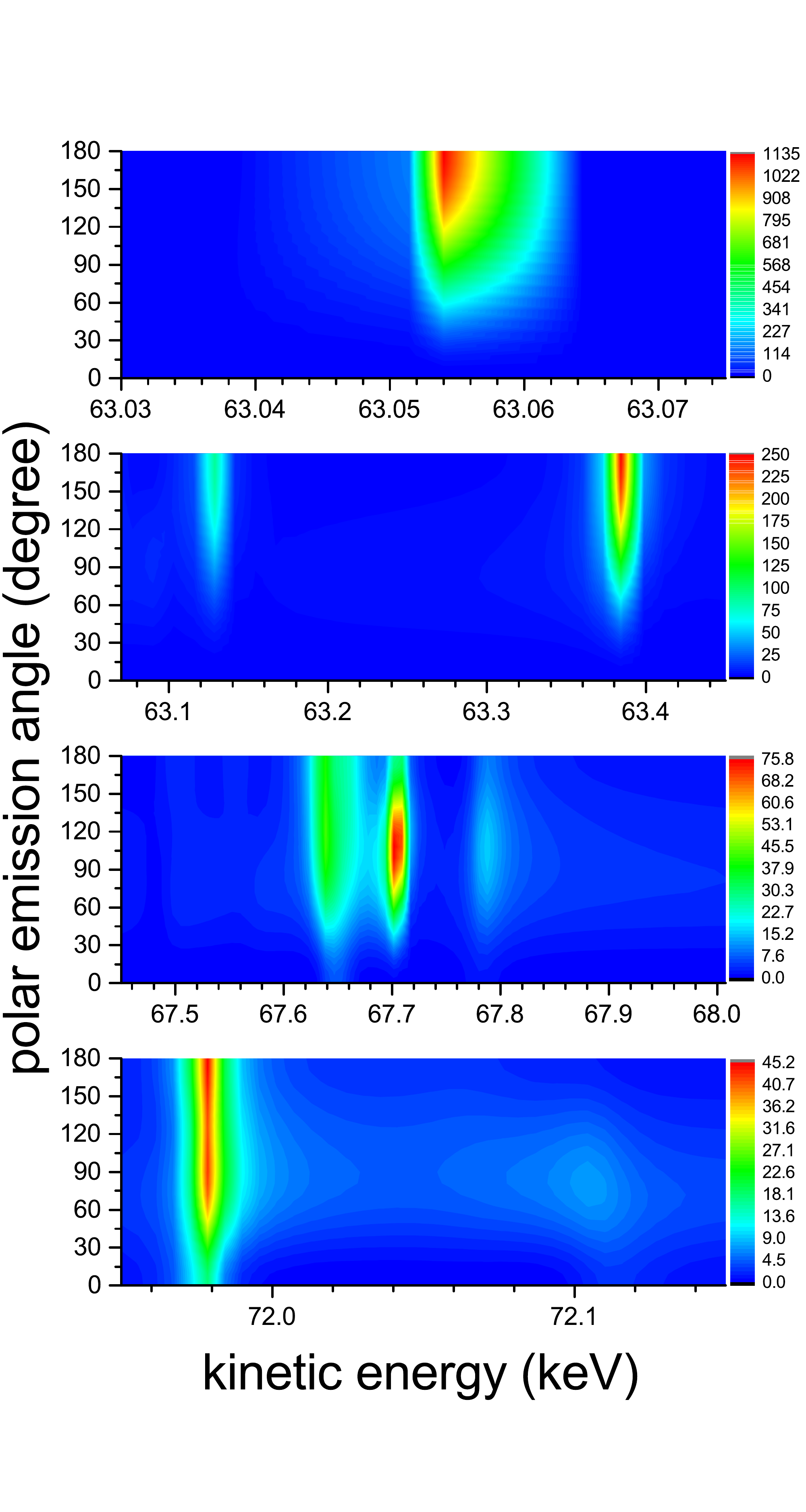}
\caption{The differential cross section ($d\sigma/d\Omega$, in barn/str) of dielectronic recombination with two-electron uranium for polarized incident electron ($\mu=-1/2$) and the photon emission with circular polarization ${\bf{e}}_{+}$ (see
\Eq{cirpol}) are presented. 
}
\label{fig10}
\end{minipage}\hspace{2pc}%
\end{figure}

We have also studied the contribution of the various polarizations of the incident electron and emitted photon.
Polarization of the initial state is defined by polarization of the incident electron (projection of the spin to the direction of momentum) which can be equal to $\mu=\pm 1/2$.
Different polarizations of the incident electron give equal contributions to the cross section, while the summation over the photon polarizations is performed; however, they give different contributions if the polarization of the emitted photon is fixed.
In
Fig. \ref{fig10}
we present results of calculation of the differential cross section $\sigma'_{-+}$, where
the incident electron has polarization $\mu=-1/2$ and emitted photon has polarization $\zhe_{+}$ (see \Eq{cirpol}).
The result of the calculation of the differential cross section $\sigma'_{--}$ (the polarization of incident electron is $\mu=-1/2$, and the photon polarization is $\zhe_{-}$) can be obtained from
Fig. \ref{fig10}
by inversion of the polar axis.
Differential cross sections with different polarizations are connected by the following condition
\begin{eqnarray}
\sigma'_{\mu,-}
&=&
\sigma'_{-\mu, +}
\,.
\end{eqnarray}

\par
\begin{figure}[h]
\begin{minipage}{40pc}
\includegraphics[width=40pc]{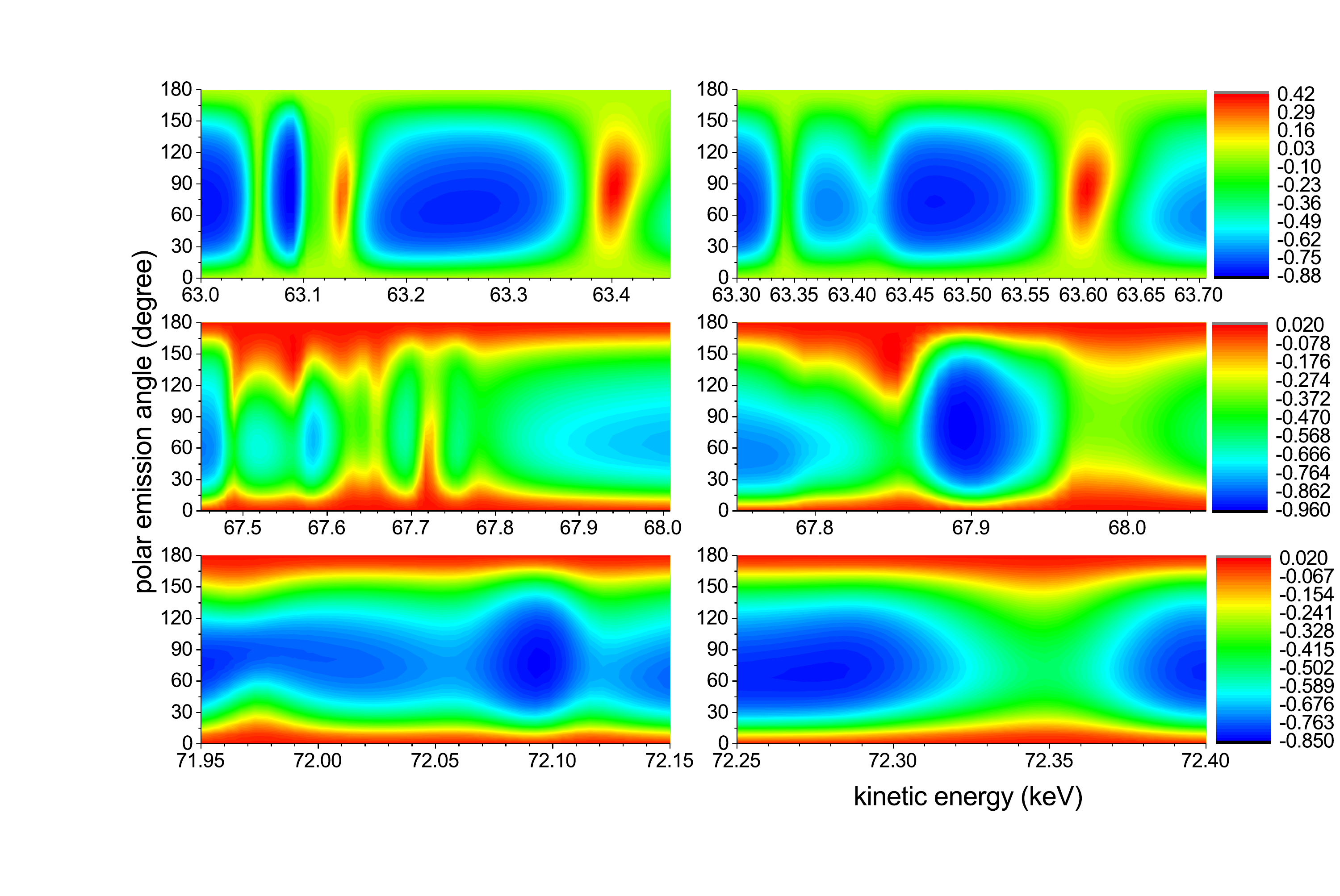}
\caption{The Stokes parameter describing the linear polarization
$P_1$ (see
\Eq{sp1})
is presented.
The graphs in the left column correspond to the full QED calculation (Coulomb + Breit with retardation) of the differential cross section, the graphs in the right column present the calculation with disregard of the Breit interaction.}
\label{fig5}
\end{minipage}\hspace{2pc}%
\end{figure}
\begin{figure}[h]
\begin{minipage}{40pc}
\includegraphics[width=40pc]{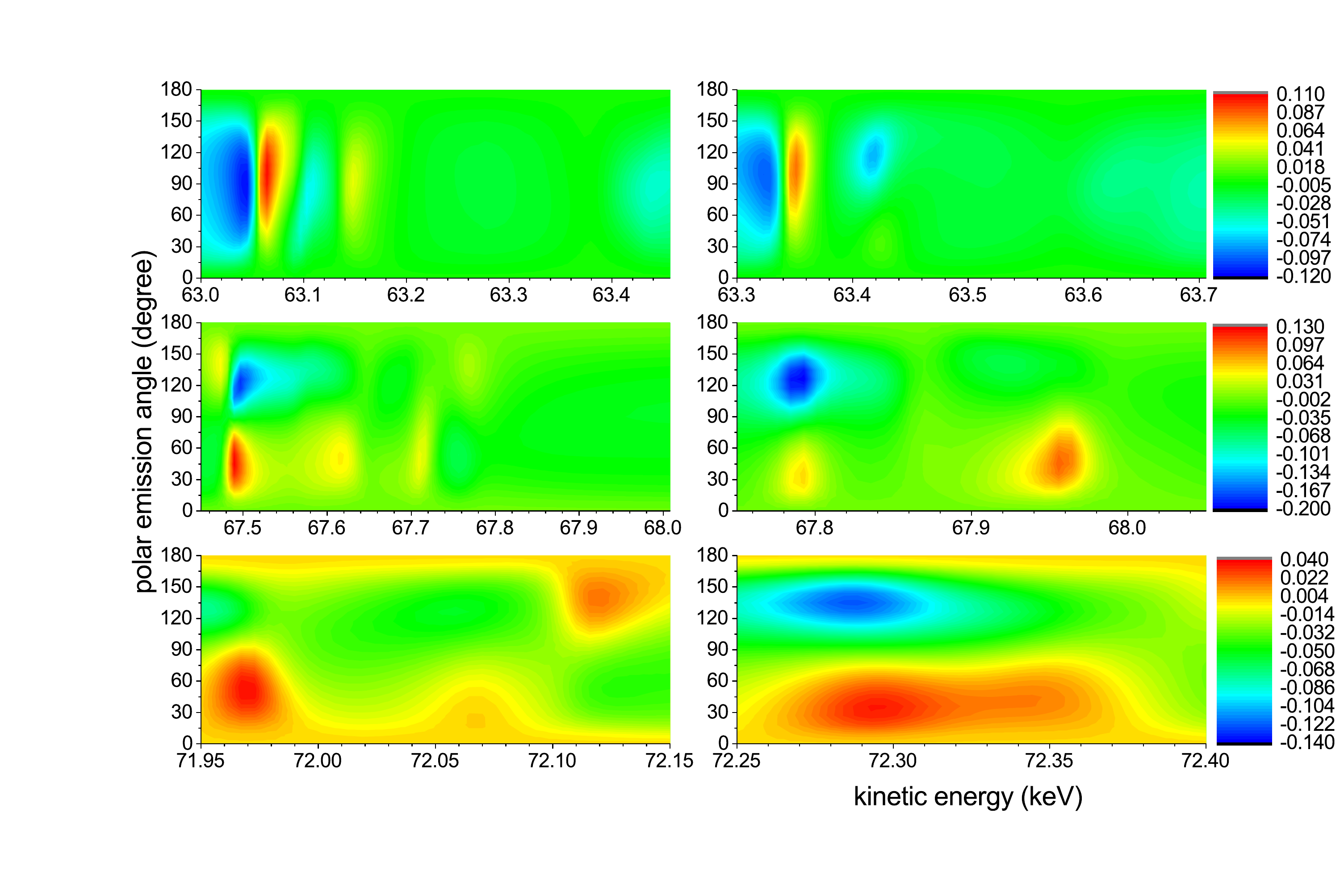}
\caption{The Stokes parameter describing the linear polarization
$P_2$ (see
\Eq{sp2})
is presented.
The graphs in the left column correspond to the full QED calculation (Coulomb + Breit with retardation) of the differential cross section, the graphs in the right column present the calculation with disregard of the Breit interaction.}
\label{fig6}
\end{minipage}\hspace{2pc}
\end{figure}
\begin{figure}[h]
\begin{minipage}{40pc}
\includegraphics[width=40pc]{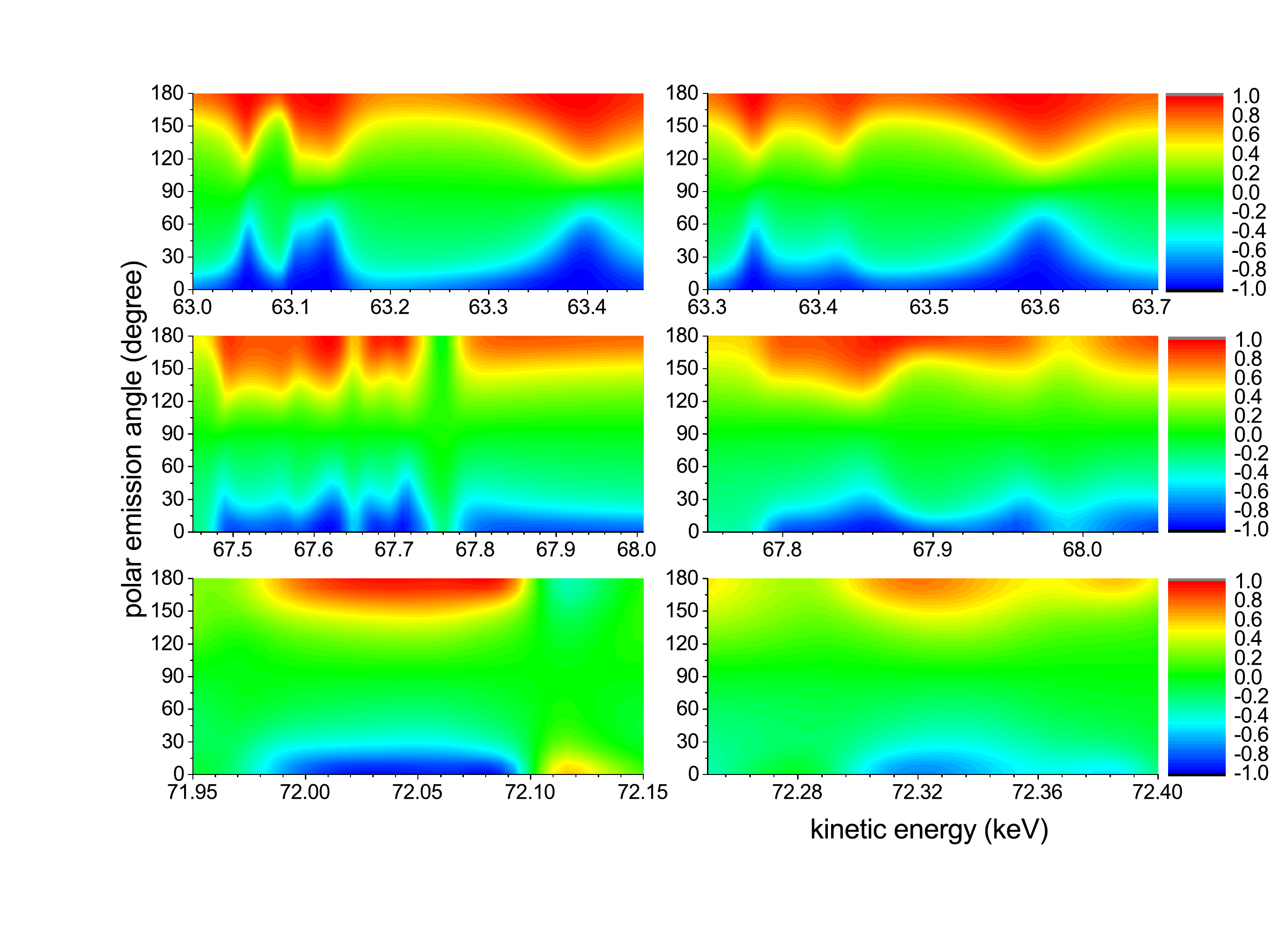}
\caption{The Stokes parameter describing the circular polarization $P_3$ (see
\Eq{sp3})
is presented.
The graphs in the left column correspond to the full QED calculation (Coulomb + Breit with retardation) of the differential cross section, the graphs in the right column present the calculation with disregard of the Breit interaction.}
\label{fig7}
\end{minipage}\hspace{2pc}%
\end{figure}

To investigate the polarizations of emitted photon, we calculated the Stokes parameters.
We have calculated the Stokes parameters for incident electrons with polarization $\mu=-1/2$; the results of the calculations are presented in
Figs. \ref{fig5}-\ref{fig7}.
The Stokes parameters $P_1$, and $P_2$ for different linear polarizations of the photon are given in
Figs. \ref{fig5}, \ref{fig6}
\begin{eqnarray}
P_1
&=&\label{sp1}
\frac{\sigma'_{0^{\circ}}-\sigma'_{90^{\circ}}}
{\sigma'_{0^{\circ}}+\sigma'_{90^{\circ}}}
\,,
\\
P_2
&=&\label{sp2}
\frac{\sigma'_{45^{\circ}}-\sigma'_{135^{\circ}}}{\sigma'_{45^{\circ}}+\sigma'_{135^{\circ}}}
\,,
\end{eqnarray}
where $\sigma'_{0^{\circ}}$, $\sigma'_{90^{\circ}}$ are the differential cross section for emission of photon with polarization vector laid or orthogonal to the ($\zhp,\zhk$) plane, respectively, and
$\sigma'_{45^{\circ}}$, $\sigma'_{135^{\circ}}$ are 
the differential cross sections for emission of photon with polarization vector at $45^{\circ}$ and $135^{\circ}$ to the ($\zhp,\zhk$) plane, respectively.
$P_1$ and $P_2$ equal to zero at angles $0^{\circ}$ and $180^{\circ}$ (see \Eq{pol1}). 
The Stokes parameter ($P_3$) describing the circular polarization
\Eq{cirpol}
is presented in
Fig. (\ref{fig7})
\begin{eqnarray}
P_3
&=&\label{sp3}
\frac{\sigma'_{+}-\sigma'_{-}}{\sigma'_{+}+\sigma'_{-}}
\,,
\end{eqnarray}
where $\sigma'_{\pm}\equiv d\sigma_{\pm}/d\Omega$ are the differential cross section for emission of the photon with the corresponding chirality.
Figs.  \ref{fig5}-\ref{fig7}
show that Stokes parameters also very sensitive to Breit interaction.

For nonpolarized
incident electrons the corresponding polarizations of the emitted photon
give equal contributions to the differential cross sections
$\sigma'_{45^{\circ}}$, $\sigma'_{135^{\circ}}$ and $\sigma'_{-}$, $\sigma'_{+}$, respectively.
Accordingly, the Stokes parameters $P_2$ and $P_3$ are equal to zero.
The parameter $P_1$ is independent of the polarization of the incident electron and
is the same for the polarized and nonpolarized incident electrons.

\par
\begin{figure}[h]
\begin{minipage}{40pc}
\includegraphics[width=40pc]{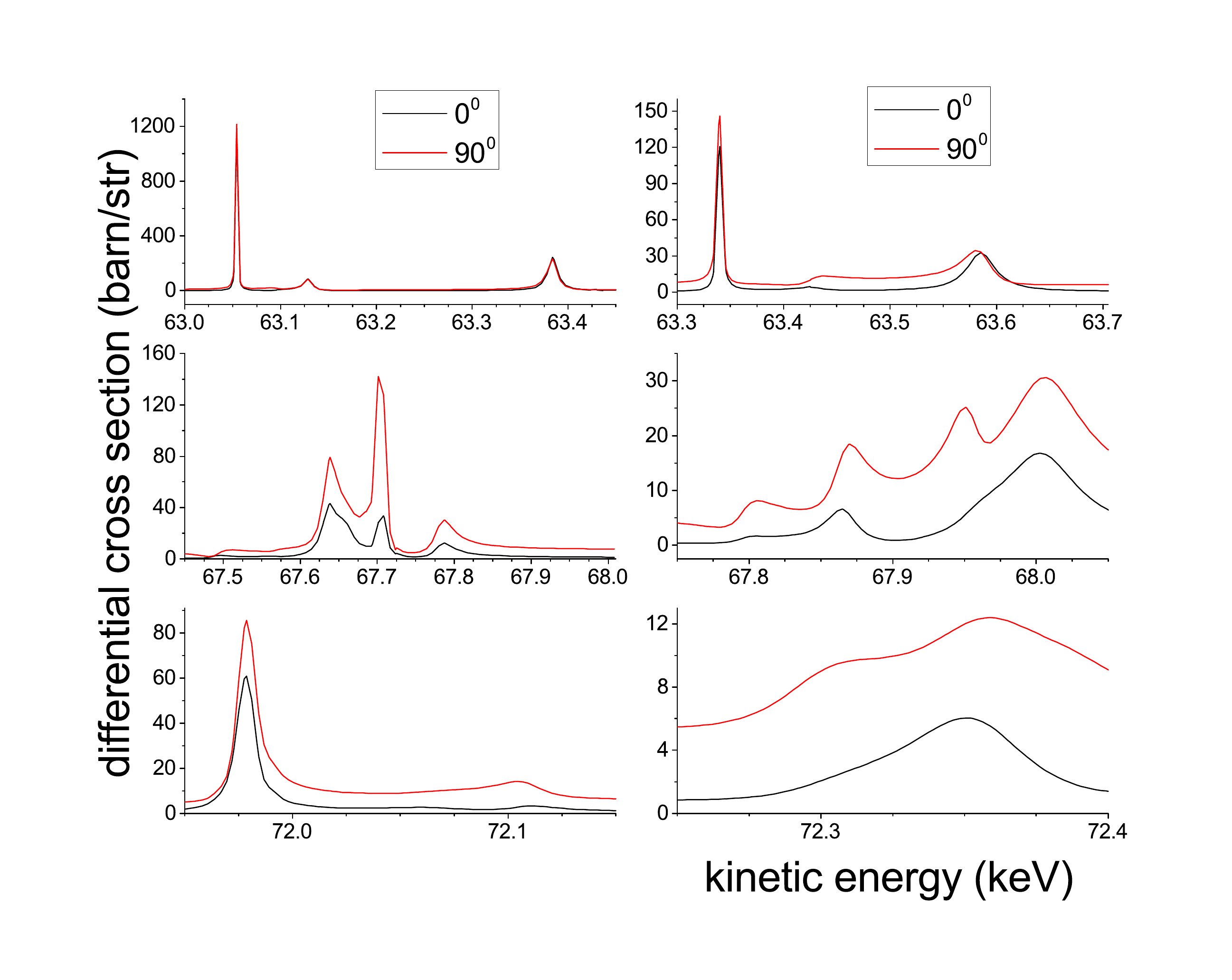}
\caption{The comparison of the differential cross section  for photon emission at $0^{\circ}$ angle ($\sigma'(\theta=0^{\circ})$, black curve, in barn/str) and the differential cross section for photon emission at $90^{\circ}$ angle ($\sigma'(\theta=90^{\circ})$, red curve, in barn/str).
The graphs in the left column correspond to the full QED calculation (Coulomb + Breit with retardation) of the differential cross section, the graphs in the right column present the calculation with disregard of the Breit interaction.
}
\label{fig8}
\end{minipage}\hspace{2pc}%
\end{figure}
\begin{figure}[h]
\begin{minipage}{40pc}
\includegraphics[width=40pc]{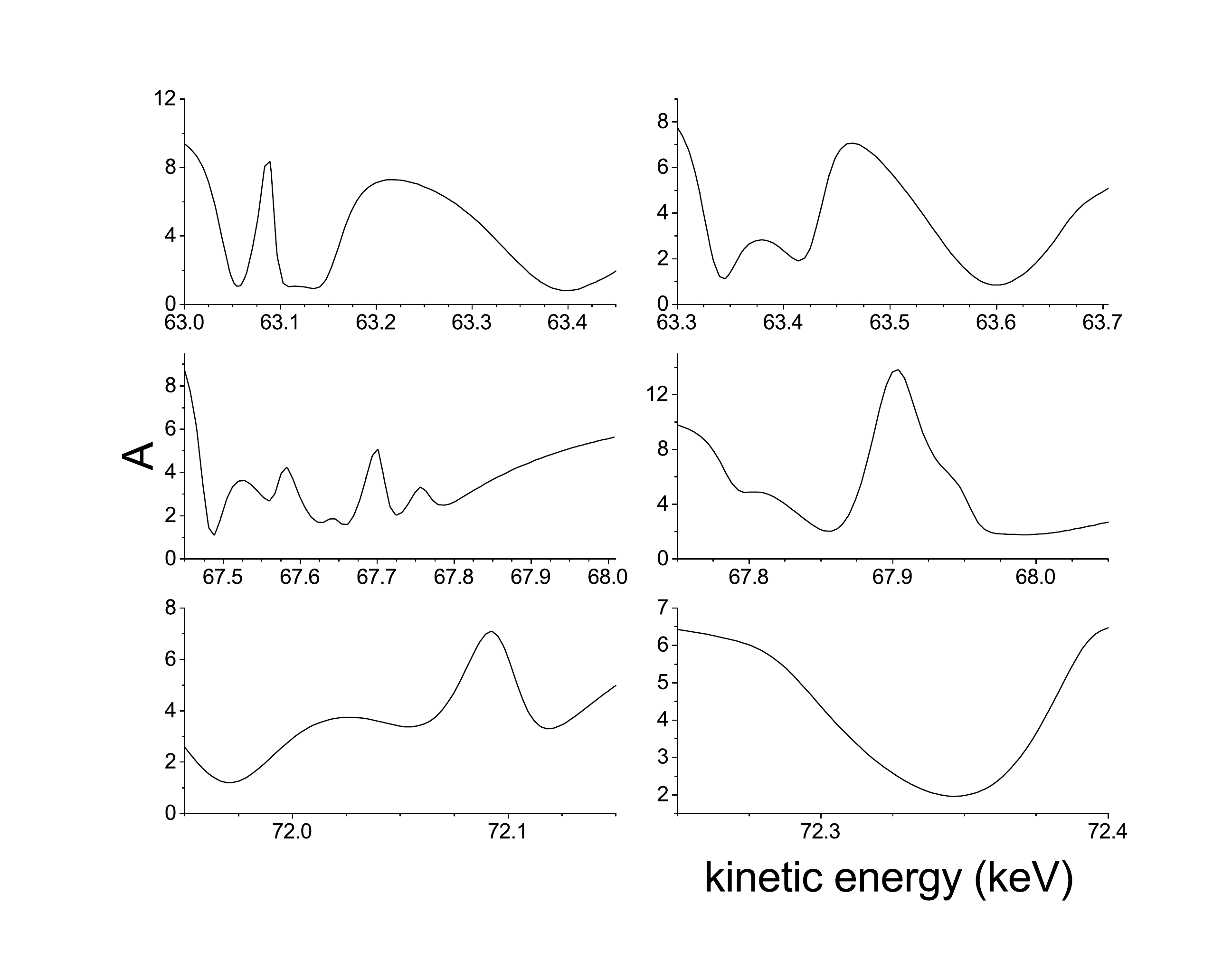}
\caption{The asymmetry parameter $A$
(see
\Eq{asym})
is presented.The graphs in the left column correspond to the full QED calculation (Coulomb + Breit with retardation) of the differential cross section, the graphs in the right column present the calculation with disregard of the Breit interaction.
}
\label{fig9}
\end{minipage}\hspace{2pc}%
\end{figure}

For addition characteristics of angular distribution we present the differential cross section with photon emission angles
$0^{\circ}$ ($\sigma'(\theta = 0^{\circ})$),
$90^{\circ}$ ($\sigma'(\theta = 90^{\circ})$)
in
Fig. \ref{fig8}
and asymmetry parameter
\cite{shah2014hci}
\begin{eqnarray}
A
&=&\label{asym}
\frac{\sigma'(\theta = 90^{\circ})}{\sigma'(\theta = 0^{\circ})}
\end{eqnarray}
in
Fig. (\ref{fig9}),
respectively.
The calculations show that the emission to $90^{\circ}$ dominates over emission to $0^{\circ}$, particularly in regions of the resonance energies.
Far from the resonance regions the photon is emitted mainly to $90^{\circ}$ angle what results in corresponding growth of the parameter $A$.

We would like to note that we have also investigated the contribution of tri-electronic recombination to the process of electron capture by two-electron uranium ion initially being in its ground state.
We performed calculations of the cross section for regions of the incident electron energy where the contribution of the triply excited states
$((2s 2s)_0 2p_{1/2})_{1/2}$ and $((2p_{1/2} 2p_{1/2})_0 2s)_{1/2}$
could be significant.
It was found that the contributions of these states to the cross section is in $8$-$11$ orders smaller than the corresponding contribution of the (nonresonant) radiative electron capture (REC), and it offers no possibility to detect the tri-electronc recombination in  experiment.
However, we found that for the process of electron capture with two-electron uranium initially being in a single excited state (for example, $(1s2s)_0$),
the contribution of the tri-electronic recombination is much larger than the corresponding contribution of the radiative electron capture, which, in principle, makes it possible to experimentally investigate the trielectronic recombination.

We have presented QED calculations of the total and differential cross section for dielectronic recombination of nonpolarized and polarized electrons with two-electron uranium initially being in its ground state.
We have also investigated contribution of the higher multipoles of the photon wave-function expansion and found that their contributions to the deferential cross section are significant.
The polarization of the emitted photons and the photon emission asymmetry are investigated.
It was found that contribution of the Breit interaction is very important for the total and differential cross section as well as for various polarization parameters.
The contribution of the Breit interaction is very significant, which allows to perform
a successful experimental investigation of the Breit
interaction in the process of dielectronic recombination with
two-electron uranium.

\begin{table}
  \caption{Comparison of the energies ($E-3mc^2=\Delta E - i\frac{\Gamma}{2}$, in keV) of three electronic states and corresponding resonance kinetic energies ($\epsilon$) of the impact electron (on the second line for each state).
Second and third columns show the diagonal element of the matrix $V$ (see \Eq{V_mat}).
Columns 4-6 show  eigenvalues of the matrix $V$
with different electron-electron interaction models. }
\label{table2}
\begin{center}
\begin{tabular}{c|cc|cc|cc|cc|cc}
\hline
& \multicolumn{2}{|c|} {V}&\multicolumn{2}{|c|} {V}& \multicolumn{2}{|c|} {}&\multicolumn{2}{|c|} {} & \multicolumn{2}{|c} {} \\

& \multicolumn{2}{|c|} {Coulomb}&\multicolumn{2}{|c|} {Coulomb+Breit}& \multicolumn{2}{|c|} {Coulomb}&\multicolumn{2}{|c|} { Coulomb+Breit} & \multicolumn{2}{|c} {Coulomb+Breit} \\

 & \multicolumn{2}{|c|} {}& \multicolumn{2}{|c|} {(with retardation) }& \multicolumn{2}{|c|}{}  &\multicolumn{2}{|c|}{(without retardation)} & \multicolumn{2}{|c}{(with retardation)}\\[5pt]
\hline
  Three electron   &$\Delta E$ &$\Gamma$&$\Delta E$ &$\Gamma$& $\Delta E$ & $\Gamma$& $\Delta E$ & $\Gamma$&$\Delta E$ &$\Gamma$\\[5pt]
   state& \multicolumn{2}{|c|} {$\epsilon$}&\multicolumn{2}{|c|} {$\epsilon$}& \multicolumn{2}{|c|} {$\epsilon$}&\multicolumn{2}{|c|} {$\epsilon$} & \multicolumn{2}{|c} {$\epsilon$} \\

\hline
\ttfamily $(1s(2s \, 2s)_0)_{1/2}$ &-198.350 & 0.000&-198.304 & 0.000&-198.377 & 0.004&-198.327 & 0.004&-198.332 & 0.002\\
 &\multicolumn{2}{|c|}{63.366}&\multicolumn{2}{|c|}{63.078}&\multicolumn{2}{|c|}{63.340}&\multicolumn{2}{|c|}{63.058}&\multicolumn{2}{|c}{63.054}\\

\hline
\ttfamily $(1s(2s \, 2p_{1/2})_1)_{3/2}$ &-198.282 & 0.031&-198.281 & 0.021&-198.291 & 0.031&-198.281 & 0.031&-198.289 & 0.021 \\
 &\multicolumn{2}{|c|}{63.433}&\multicolumn{2}{|c|}{63.102}&\multicolumn{2}{|c|}{63.426}&\multicolumn{2}{|c|}{63.105}&\multicolumn{2}{|c}{63.097}\\
\hline

\ttfamily $(1s(2s \, 2p_{1/2})_0)_{1/2}$ &-198.299 & 0.032&-198.231 & 0.016&-198.357 & 0.032& -198.245 & 0.032&-198.256 & 0.011\\
 &\multicolumn{2}{|c|}{63.416}&\multicolumn{2}{|c|}{63.151}&\multicolumn{2}{|c|}{63.360}&\multicolumn{2}{|c|}{63.140}&\multicolumn{2}{|c}{63.129}\\
\hline

\ttfamily $(1s(2s \, 2p_{1/2})_1)_{1/2}$ &-198.174 & 0.032&-198.009 & 0.006&-198.131 & 0.032&-198.000 & 0.032&-198.002 & 0.012\\
 &\multicolumn{2}{|c|}{63.541}&\multicolumn{2}{|c|}{63.373}&\multicolumn{2}{|c|}{63.585}&\multicolumn{2}{|c|}{63.385}&\multicolumn{2}{|c}{63.384}\\
\hline

\ttfamily $(1s(2p_{1/2} \, 2p_{1/2})_0)_{1/2}$ &-198.058 & 0.063&-197.962 & 0.032&-198.048 & 0.059&-197.944 & 0.060&-197.954 & 0.031\\
 &\multicolumn{2}{|c|}{63.658}&\multicolumn{2}{|c|}{63.421}&\multicolumn{2}{|c|}{63.669}&\multicolumn{2}{|c|}{63.441}&\multicolumn{2}{|c}{63.431}\\
\hline

\ttfamily $(1s(2s \, 2p_{3/2})_1)_{3/2}$ &-193.832 & 0.027&-193.795 & 0.003&-193.769 & 0.027&-193.679 & 0.027&-193.901 & 0.026\\
 &\multicolumn{2}{|c|}{67.883}&\multicolumn{2}{|c|}{67.588}&\multicolumn{2}{|c|}{67.948}&\multicolumn{2}{|c|}{67.706}&\multicolumn{2}{|c}{67.485}\\
\hline

\ttfamily $(1s(2s \, 2p_{3/2})_1)_{1/2}$ &-193.856 & 0.026&-193.821 & 0.034&-193.857 & 0.026&-193.824 & 0.026&-193.822 & 0.034\\
 &\multicolumn{2}{|c|}{67.859}&\multicolumn{2}{|c|}{67.561}&\multicolumn{2}{|c|}{67.860}&\multicolumn{2}{|c|}{67.561}&\multicolumn{2}{|c}{67.563}\\
\hline

\ttfamily $(1s(2p_{1/2} \, 2p_{3/2})_2)_{5/2}$ &-193.745 & 0.058&-193.747 & 0.022&-193.750 & 0.058&-193.730 & 0.058&-193.752 & 0.022\\
 &\multicolumn{2}{|c|}{67.970}&\multicolumn{2}{|c|}{67.635}&\multicolumn{2}{|c|}{67.967}&\multicolumn{2}{|c|}{67.655}&\multicolumn{2}{|c}{67.634}\\
\hline

\ttfamily $(1s(2p_{1/2} \, 2p_{3/2})_1)_{3/2}$ &-193.788 & 0.058&-193.717 & 0.016&-193.801 & 0.058&-193.718 & 0.058&-193.732 & 0.029\\
 &\multicolumn{2}{|c|}{67.928}&\multicolumn{2}{|c|}{67.665}&\multicolumn{2}{|c|}{67.915}&\multicolumn{2}{|c|}{67.667}&\multicolumn{2}{|c}{67.654}\\
\hline

\ttfamily $(1s(2p_{1/2} \, 2p_{3/2})_1)_{1/2}$ &-193.732 & 0.057&-193.690 & 0.056&-193.736 & 0.057&-193.698 & 0.057 &-193.695 & 0.056\\
 &\multicolumn{2}{|c|}{67.983}&\multicolumn{2}{|c|}{67.692}&\multicolumn{2}{|c|}{67.980}&\multicolumn{2}{|c|}{67.687}&\multicolumn{2}{|c}{67.691}\\
\hline

\ttfamily $(1s(2s \, 2p_{3/2})_2)_{3/2}$ &-193.860 & 0.026&-193.788 & 0.032&-193.927 & 0.026&-193.897 & 0.026&-193.686 & 0.009\\
 &\multicolumn{2}{|c|}{67.855}&\multicolumn{2}{|c|}{67.594}&\multicolumn{2}{|c|}{67.790}&\multicolumn{2}{|c|}{67.488}&\multicolumn{2}{|c}{67.700}\\
\hline

\ttfamily $(1s(2p_{1/2} \, 2p_{3/2})_2)_{3/2}$ &-193.723 & 0.058&-193.610 & 0.041&-193.717 & 0.058&-193.594 & 0.058&-193.606 & 0.029\\
 &\multicolumn{2}{|c|}{67.993}&\multicolumn{2}{|c|}{67.772}&\multicolumn{2}{|c|}{67.999}&\multicolumn{2}{|c|}{67.791}&\multicolumn{2}{|c}{67.780}\\
\hline

\ttfamily $(1s(2p_{3/2} \, 2p_{3/2})_2)_{5/2}$ &-189.426 & 0.053&-189.415 & 0.010&-189.426 & 0.053&-189.385 & 0.053&-189.414 & 0.010\\
 &\multicolumn{2}{|c|}{72.289}&\multicolumn{2}{|c|}{71.968}&\multicolumn{2}{|c|}{72.291}&\multicolumn{2}{|c|}{72.001}&\multicolumn{2}{|c}{71.971}\\
\hline

\ttfamily $(1s(2p_{3/2} \, 2p_{3/2})_2)_{3/2}$ &-189.369 & 0.052&-189.324 & 0.052&-189.369 & 0.052&-189.325 & 0.052&-189.324 & 0.052\\
 &\multicolumn{2}{|c|}{72.346}&\multicolumn{2}{|c|}{72.058}&\multicolumn{2}{|c|}{72.348}&\multicolumn{2}{|c|}{72.060}&\multicolumn{2}{|c}{72.061}\\
\hline

\ttfamily $(1s(2p_{3/2} \, 2p_{3/2})_0)_{1/2}$ &-189.334 & 0.053&-189.288 & 0.027&-189.332 & 0.053&-189.269 & 0.053&-189.286 & 0.027\\
 &\multicolumn{2}{|c|}{72.381}&\multicolumn{2}{|c|}{72.094}&\multicolumn{2}{|c|}{72.385}&\multicolumn{2}{|c|}{72.117}&\multicolumn{2}{|c}{72.100}\\

\hline

\end{tabular}
\end{center}
\end{table}

\acknowledgments
The authors are indebted to senior lecture V.V. Bankevich for reading and improving the manuscript.
The authors acknowledge the support by RFBR grants 14-02-20188 and
the support by
St. Petersburg State University through a research grant
11.38.227.2014.
The work of K.N.L. was supported by RFBR grant 16-32-00620 and G-RISC P-2016a-8.


%
%

%
%

\end{document}